\documentstyle[11pt,psfig,aaspp4] {article}
\begin{document}
\title{The X-ray Iron Emission from Tycho's Supernova Remnant}
\author{Una Hwang\altaffilmark{1,2}, John P. Hughes\altaffilmark{3}, \& Robert Petre\altaffilmark{1}}
\altaffiltext{1}{NASA Goddard Space Flight Center, Laboratory for
High Energy Astrophysics, Greenbelt, MD 20771}
\altaffiltext{2}{Department of Astronomy, University of Maryland at
College Park, MD 20742}
\altaffiltext{3}{Department of Physics and Astronomy, Rutgers
University, P.O. Box 849, Piscataway, NJ 08855}

\def\etal{{\it et~al.}}
\def\msun{M$_\odot$}

\begin{abstract}

We present the results of broadband spectral model fits to the X-ray
spectrum of Tycho's supernova remnant obtained by the Solid-State
Imaging Spectrometers on the ASCA Observatory.  We use
single-temperature, single-ionization-age, nonequilibrium ionization
models to characterize the ejecta and the blast-shocked interstellar
medium.  Previous spectral studies have suggested that the Fe ejecta
in this Type Ia remnant are stratified interior to the other ejecta.
These studies, however, have generally restricted their attention to
the Fe K blend at 6.5 keV.  ASCA provides data over a broad energy
range that allow the placement of important constraints simultaneously
from the Fe L emission near 1 keV as well as the Fe K emission.  We
find that the simplest models, with emission from the ejecta and blast
wave each at a single temperature and ionization age, severely
underestimate the flux in the Fe K blend.  These models also show that
there is little Fe emission associated with the Si and S ejecta shell,
which has an average temperature $kT \sim$ 0.86 keV and ionization age
$n_et \sim 10^{11}\ \rm{cm}^{-3}$ s.  The blast-shocked interstellar
medium has abundances roughly 0.3 times the solar value, while the
ejecta, with the exception of Fe, have relative abundances that are
typical of Type Ia supernovae.  The addition of another component of
Fe emission, which we associate with ejecta, at a temperature at least
two times higher and an ionization age $\sim$ 100 times lower than the
Si ejecta, does provide a good fit to the spectrum.  This model is
consistent with imaging results.  Although fluorescent emission from
dust in the remnant may contribute to the Fe K flux, we conclude that
it is unlikely to be the dominant component.

\end{abstract}

\newpage
\section{Introduction}

An important issue concerning supernova ejecta is whether or not the
radially distributed layers formed during hydrostatic and explosive
nucleosynthesis are preserved, or if these layers are mixed together.
According to models of Type Ia supernovae (explosive accretion-induced
burning instabilities in white dwarfs), the ejecta are expected to
retain some of their stratification (e.g., Nomoto et al. 1984).
Although the outer layers of the ejecta may be mixed by convective
instabilities, as is suggested by the early-time spectra of Type Ia
supernovae (Branch et al. 1985), the Fe ejecta may remain distinct
from other ejecta in an inner layer, and thus be only partially heated
by the reverse shock.  This scenario is supported by UV observations
of the supernova remnant SN1006 (Wu et al. 1983, 1993, 1997).
Absorption due to cold Fe is observed in the continuum emission from a
background star, providing evidence for the presence of unshocked Fe
ejecta in the interior of the remnant.  By contrast, observations of
SN 1987a show that the ejecta are evidently well-mixed in the remnants
of supernovae caused by the gravitational collapse of massive
progenitors (Type II, Ib, and Ic; Woosley 1991 and references
therein).

The remnant of the supernova recorded by Tycho Brahe in 1572 is
generally considered to be the prototype of Type Ia remnants.  It has
been studied extensively at all wavelengths, but information about its
ejecta can now be obtained only from its X-ray emission.  The optical
emission from this young remnant is virtually all H Balmer emission
due to the excitation of partially neutral ambient material at the
outer shock wave (Smith et al. 1991).  Even in remnants with bright
optical emission from the ejecta, the mass of optically emitting
material represents only a tiny fraction of the total ejecta mass;
X-rays are the dominant emission at the characteristic temperature of
the shock-heated material which comprises a supernova remnant (SNR).
Hamilton, Sarazin, \& Szymkowiak (1986) were the first to suggest that
much of the Fe ejecta could be hidden in an inner layer.  They applied
a self-similar hydrodynamical model for the ejecta in Tycho to a
number of observations over a wide range of energies.  In their model,
a small amount of Fe ejecta mixed together with Si and other ejecta
provide most of the X-ray emission from Fe; the inner layers of Fe are
too cool to provide significant X-ray emission.  A number of other
spectral studies suggest that the Fe ejecta are stratified interior to
the lighter ejecta.  Hughes (1991) and Petre et al. (1992) showed that
the centroid of the Fe K blend requires a significantly lower
ionization age than the other elements.  The ionization age is defined
as $n_e t$, the product of electron density with time since the gas
was shock-heated.  Because remnants are low-density plasmas (electron
density $n_e \sim 1\ \rm{cm}^{-3}$), the time required to attain the
equilibrium population distribution of ions at a given temperature
($\sim 10^4/n_e$ yr) generally exceeds the age of the remnant.  The
low ionization age of the Fe emission could be interpreted as evidence
for the stratification of Fe ejecta within interior layers of the
remnant that have only recently been heated by the reverse shock.
However, another possibility is that the Fe K emission arises in lower
density ambient material shocked by the blast wave.

More recent observations with the ASCA Solid State Imaging
Spectrometers (SIS) have now provided a superior moderate resolution
spectrum of Tycho's SNR at energies between 0.4$-$10 keV with
$1\arcmin$ spatial resolution.  Hwang \& Gotthelf (1997; hereafter
HG97) use these data in a study of the line emission.  Using measured
emission line intensity ratios and centroids of emission line blends,
they conclude with other authors that the Fe emission arises under
conditions different from Si and S, namely a higher temperature and
lower ionization age.  Their narrowband ASCA images also show that the
spatial distributions of Fe L and Fe K emission are not the same.  The
Fe K emission is somewhat more centrally distributed than all other
X-ray spectral features distinguishable by ASCA.  This central
distribution of the Fe K emission suggests that it may arise from hot
inner ejecta layers and not from the blast wave.

As suggestive as this evidence is, the nature of the Fe emission in
Tycho's SNR merits a more careful inquiry.  A difference between the
average spectral properties of the elements Si and Fe does not
necessarily require that they be spatially separate.  A range of
temperatures coupled with the corresponding differences in emissivity
for different ions could allow them to occupy the same volume, with
the Si emission dominating at the lower temperatures and the Fe
emission at higher temperatures.  The X-ray spectrum is also clearly
complex, and emission from the blast-shocked interstellar medium is
expected to make a contribution.  High resolution images from the
Einstein Observatory (Seward, Gorenstein, \& Tucker 1983) resolve a
thin shelf of emission attributed to the blast wave, which is
estimated to contribute 28\% of the counts in the Einstein energy
bandpass (0.5$-$4.5 keV).  Instruments with sensitivity at higher
energies have also clearly detected harder X-ray emission (Pravdo \&
Smith 1979, Fink et al. 1994).  Model fitting is therefore necessary.

With the exception of the hydrodynamic similarity model of HSS86, the
results presented so far in support of the stratification of the Fe
ejecta are based on the line diagnostics approach of comparing
measured emission line intensities and centroids to theoretical
models.  They have focused on the centroid of the Fe K blend, assuming
that this emission is from the ejecta, and at best make rough
comparisons of the Fe L and Fe K intensities (Vancura, Gorenstein, \&
Hughes 1995 and HG97).  This is because the Fe L emission at energies
near 1 keV is inaccessible to line diagnostics with data obtained by
ASCA and all previous X-ray broadband spectrometers.  The numerous
line transitions that are collectively referred to as Fe L are blended
together by the ASCA spectrometers, and individual Fe L line
intensities cannot be measured relative to the continuum.  Fitting a
model to the broadband spectrum allows us to maximize our use of the
spectral data, particularly the Fe L emission.

In this paper we use the ASCA SIS spectral data to investigate further
the nature of the Fe emission by carrying out broadband model fits
including nonequilibrium ionization (NEI) effects.  We use
single-temperature, single-ionization age NEI models to represent both
the ejecta and the blast-shocked interstellar medium.  We build on the
results of HG97 by using their results for the temperature of the Si
ejecta.  The shape and intensity of the Fe L emission at energies near
1 keV provide critical constraints on these models.

\section{Procedure and Results}

The characteristics of the ASCA SIS and our data reduction procedures
are described in HG97 and the references therein.  Briefly, the SIS
consists of two slightly misaligned square arrays of 4 CCD chips, each
with its own telescope.  The telescope provides imaging with a narrow
1$'$ core and extended wings with a half-power diameter of 3$'$.  Each
SIS can be operated with 1, 2, or all 4 of its chips exposed.  Spectra
are obtained over energies of 0.4$-$10 keV with a spectral resolution
of 2\% at 6 keV (resolution at launch) that scales with energy like
E$^{-1/2}$.  The data for Tycho's SNR were taken in 2- and 4-CCD mode
on 29 August 1993 during the Performance Verification phase of the
mission.  In this paper, we use only the spectral data taken in 2-CCD
mode to minimize the impact of calibration uncertainties, which are
significantly greater for data taken in 4-CCD mode.  Because Tycho is
so bright in X-rays, its spectrum is particularly sensitive to such
uncertainties.  We study in detail the spectrum obtained by the
best-calibrated CCD chip (chip 1 on SIS0), which covers roughly the
southeastern quadrant of the remnant.  The ASCA spectrum of Tycho's
SNR shows only relatively modest variations with position (HG97).

Line widths are clearly in excess of the predicted instrumental
widths.  We find that a Gaussian broadening corresponding to a
velocity of about 7500 km s$^{-1}$ FWHM is adequate for all the
emission lines.  This cannot be considered a measurement of physical
velocities in the remnant: the instrumental response is not yet
perfectly understood, and some percentage of this width is probably
attributable to imperfect calibration.

\subsection{Spectral Fits}

In our models we include the emission from the blast-shocked
interstellar medium (ISM) and the reverse-shocked ejecta, representing
each by a single-temperature, single-ionization-age, NEI component.
Although this approach sounds simple, Hamilton \& Sarazin (1984) have
shown that many commonly-invoked hydrodynamic scenarios relevant to
supernova remnants can be well-approximated with an emission-averaged
temperature and ionization parameter.  To first order, the reverse
shock and blast shock should each be characterizable in this way.

Such a model, however, cannot be a perfect representation of an object
as complex as a young supernova remnant.  The X-ray emitting gas in
remnants has a distribution of temperatures and densities arising from
their complex hydrodynamic evolution.  Our goals are to characterize
as simply as possible the emission-weighted average properties of the
important contributions to the X-ray spectrum.  With this approach,
one is not constrained {\it a priori} to an explicit hydrodynamic
model.  An important advantage is that the reduced computational
burden allows us to explore a wider range of parameter space.  This
approach has been applied successfully in the past to study other
supernova remnants (e.g., Hughes \& Singh 1994, Hwang et al. 1993).

The spectral fits employ the X-ray thermal emission models of Raymond
\& Smith (1977, with updates 1992) with the calculation of the
nonequilibrium ionization fractions carried out using matrix
diagonalization techniques (Hughes \& Helfand 1985).  The NEI code
uses ionization and recombination rates from the basic Raymond \&
Smith code to calculate the time-dependent ionization state of gas
which has been impulsively heated to a high temperature by the passage
of a shock wave.  Additional lines appropriate to the NEI situation
have also been added to the model using data from Mewe, Gronenschild,
\& van den Oord (1985), as explained in Hughes \& Singh (1994).  See
this reference for further discussion of NEI effects on X-ray spectra.
In our fits, the abundances of the elements are quoted relative to the
solar values of Allen (1973; Fe/H = 3.98$\times 10^{-5}$ by number).

In the following three sections, we present fits of models with one,
two, and three NEI components.

\subsection{Against a Blast Wave Origin for the Fe K Emission}

HG97 report a temperature $kT$ of about 0.86 keV and an ionization age
$n_et$ of about $10^{11}$ cm$^{-3}$ s for the Si and S ejecta based on
their analysis of the emission line intensity ratios.  These and other
authors note that there is additional emission at energies above about
5 keV.  To better characterize this hard emission, we first fitted the
spectrum at energies above 4.5 keV with a NEI component for Fe and Ni
and an additional component from H and He at the same temperature
providing most of the continuum emission; the abundances of Fe and Ni
were required to be the same.  At energies above 4.5 keV, the 0.86 keV
ejecta component contributes very little to the spectrum.

The fits were performed for a grid of temperatures $kT$ from 2.5 to 20
keV, as summarized in Table 1.  The ionization age is about $10^{10}\
{\rm cm}^{-3}$ s for the entire range of temperatures, and the
abundance required to reproduce the Fe K line intensity is 1.5$-$3.5
times the solar value; at a temperature $kT =$ 20 keV, the lower limit
for the abundance ($\Delta\chi^2$ = 2.71) is consistent with the solar
value.  The high abundances required suggest that the Fe K emission
cannot come primarily from a blast wave in the interstellar medium
unless the ISM is significantly enriched in this region.  There is no
independent evidence for such enrichment, however.  Furthermore,
extrapolation of the best-fit model ($kT$ = 4.5 keV) throughout the
entire ASCA band shows that this model cannot be correct (Figure 1).
With the H absorption column density fixed at the nominal Galactic
value of $4.5 \times 10^{21}\ {\rm cm}^{-2}$ (Albinson et al. 1986),
the Fe L emission is severly overpredicted, even in the absence of the
$kT=$ 0.86 keV ejecta component required to explain the strong Si and
S emission.  Any reasonable increase in the column density cannot make
the model consistent with the data.

We may also consider whether a Sedov-type blast wave yields the
required intensity in the Fe K blend.  The age is known to be 421 yr
at the time of the ASCA observations.  An ambient H density of about
0.3 cm$^{-3}$ is estimated by both Seward et al. (1983) and Kirshner
et al. (1987) from X-ray and optical observations, respectively.  The
width of the broad H Balmer lines in the optical give a blast wave
velocity of 1500$-$2800 km s$^{-1}$ (Smith et al. 1991).  For a
distance of 2.3 kpc to Tycho (Green 1984), the optical and radio
proper motion studies give blast wave velocities consistent with this.
The X-ray proper motion study using the ROSAT HRI by Hughes (1996)
shows two velocity components: the lower velocity is consistent with
the radio and optical values, but the higher velocity, presumably
associated with the blast wave, is about 4600 $\pm$ 1400 km s$^{-1}$
for a distance of 2.3 kpc.  For our calculation we use the optical
velocity since it is independent of distance, but note that the
optical velocities may not be representative of the average shock
velocity.  Reynolds et al. (1996) and Hughes (1996) report significant
variation in the velocity around the rim of the remnant in the radio
and in X-rays, respectively.  Our conclusions do not change if we
adopt the higher X-ray determined blast velocity.

Using the equations given in Hamilton, Sarazin, \& Chevalier (1983),
we derive the shock temperature as $kT = 2.8-$9.6 keV and the
ionization parameter log $\eta \equiv n_o^2 E$ (erg cm$^{-6}$) =
47.8$-$49.1, where $n_o$ is the ambient H density in cm$^{-3}$ and $E$
is the explosion energy in $10^{51}$ ergs.  For these values, the
predicted flux in Fe K from a solar abundance plasma is more than an
order of magnitude lower than that measured by ASCA, and the predicted
equivalent width is at least several times too low.  Therefore, in the
context of a Sedov model, the observed Fe K emission requires the
ambient Fe abundance to be several times higher than the solar value.
%There is evidence, however,
%that a Sedov model may not be a good description of this remnant in
%that the predicted explosion energy is 10$-$100 times lower than
%expected for a Type Ia explosion.  A higher ambient density would
%increase the explosion energy, but could imply more mass in the
%shocked ISM than determined by Seward et al.  A higher blast
%temperature would also result in a higher explosion energy, but would
%require a large distance to match the predicted and observed radii.
%In neither case is the Fe K flux sufficiently increased to match the
%observations.

Based on the foregoing, we conclude that while emission from the blast
wave may provide most of the continuum at energies above 5 keV, it
cannot provide the bulk of the Fe K line emission.  At the temperature
required to model the continuum, and the ionization age required to
reproduce the Fe K blend at the observed energy, this model cannot
provide sufficient flux in the Fe K emission without severely
overpredicting the Fe L emission.  We proceed, nevertheless, to fit
the data with two NEI components representing the ejecta and the blast
wave in order to gain insights concerning the shape of the overall
spectrum, the relative abundances of the elements in the ejecta, and
the abundances in the blast-shocked ISM.

\subsection{Fits for Ejecta and Blast Components}

In these fits, the ejecta component is fixed at a temperature of $kT =
0.86$ keV following HG97.  All the important elements are included in
this component, with He, C, and N in their solar abundance ratios
relative to H, the O abundance tied to that of Ne, and Fe tied to Ni.
The abundances of O, Mg, Si, S, Ar, Ca, and Fe are therefore fitted
freely, as are the ionization parameter and the absorbing interstellar
H column density.  For the blast wave component, the element
abundances are assumed to be in their solar ratios.  We fitted the
data at energies from 0.6$-$10 keV to avoid the oxygen edge in the
detector response at 0.54 keV.

All the fitted models are conspicuously deficient in emission at
energies near 0.7 and 1.2 keV.  We have also noticed this effect when
fitting other supernova remnant spectra using this NEI model and
believe that this is most likely due to incompleteness in the lines
included in the models or inaccuracies in the atomic physics used.
Liedahl, Osterheld, \& Goldstein (1995) have recalculated Fe L
emissivities for the Fe 23 and Fe 24 ions and find some significant
deficiencies in the Fe L emissivities at energies near 1.1 keV in the
current Raymond \& Smith code.  This may partially explain the deficit
in our fits at 1.2 keV; it certainly highlights the need for updated
atomic calculations.  We therefore included two gaussian emission
lines at approximately 0.726 and 1.200 keV in the models, with
equivalent widths of 200$-$300 eV and 60$-$80 eV, respectively.  We
assume that these features are Fe lines.  The Raymond \& Smith model
has features due to Fe 16 (0.725 keV) and Fe 17 (0.727 keV), and Fe 20
(1.204 keV) and Ni 24 (1.204 keV) at approximately these
energies\footnote{Between 0.70$-$0.74 keV, the model includes lines of
Ca 16 (0.708 keV), Fe 16 (0.708, 0.725 keV) and Fe 17 (0.727, 0.739
keV); between 1.200$-$1.230 keV, the model includes lines of Ne 10
(1.211 keV), Fe 20 (1.204 keV), and Ni 19 (1.228 keV), Ni 24 (1.204
keV), and Ni 25 (1.216 keV).}.

The results of the fits for a grid of blast-wave temperatures $kT$
ranging from 2.5 to 20 keV are given in Table 2.  The minimum $\chi^2$
is obtained for a blast temperature $kT=$ 3.6 keV. The fitted column
density of $5.7 \times 10^{21}\ \rm{cm}^{-2}$ is slightly higher than
the nominal Galactic value measured in the radio (Albinson et
al. 1986), and is high enough to result in significant attenuation of
the X-ray spectrum at energies below 0.8 keV.  The abundance of Fe in
the ejecta component is low, showing that there is little Fe
associated with the gas emitting the Si and S lines.  Allowing a
separate ionization age for Fe at the same temperature as Si does not
significantly alter the fitted Fe abundance.  The Fe L emission at
energies near 1 keV is very weak, and determines the Fe abundance in
the ejecta component at a temperature $kT=$ 0.86 keV and ionization
age $n_et \sim 10^{11}$ cm$^{-3}$ s; such gas is too cool to contribute
significantly to the Fe K emission.  The dominant Fe ions are Fe 19
and Fe 20.  The weakness of the Fe L emission also imposes subsolar
abundances for the blast component (see Table), resulting in a strong
deficit in the Fe K blend emission in the models as expected from the
results of the preceding section.  This is illustrated in Figure 2,
which shows the best-fit model with temperatures $kT$ = 0.86 (ejecta)
and 3.6 keV (blast), folded through the instrument response and
overlaid on the data.  The contribution to the spectrum from the
ejecta and blast components are also shown separately.

Only the Si and Fe abundances are listed in Table 2 for the $kT=$ 0.86
keV ejecta component.  In Figure 3, we compare the ratios of the
abundances of the elements O (tied to Ne), Mg, S, Ar, Ca, and Fe in
this component to that of Si, relative to the solar values of Allen
(1973), for the entire set of fits in Table 2.  The relative
abundances of the intermediate mass elements Si, S, Ar, and Ca vary by
less than 50\% over the entire range of temperatures; the abundances
of O, Mg, and Fe vary somewhat more.  Shown in the same figure are the
predicted abundances of these elements relative to Si relative to the
solar abundances of Allen (1973) in the W7 model for a Type Ia
supernova of Nomoto et al. (1984) and the updated calculations of
Thielemann et al. (1993).  Our fitted abundances for Si, S, Ar, and Ca
are in good qualitative agreement with the Nomoto et al. results; the
newer calculations predict a lower Ca mass, which makes Ca/Si about a
factor three lower than in our fits.  It should be noted here that the
high abundance of Ca relative to Fe makes the Ca L shell emission
important at energies near 0.7 keV and below.  The Ca abundance may be
constrained more by this part of the spectrum than by the Ca K blend,
but the L-shell atomic physics are probably not as reliable as that
for the K-shell (cf.  Liedahl et al. 1995).  The influence of the Ca L
shell lines may also be responsible for the line energy of the Ca K
blend at energy $\sim$ 4.9 keV appearing to be slightly too high
relative to the data.  Alternatively, the ionization age for Ca may
actually be different from that for Si and S.

The results in Table 2 also show that the fitted ionization age $n_et$
for the hotter component is near 10$^{10}$ cm$^{-3}$ s for the entire
range of temperatures considered.  The ionization age is determined
primarily by the shape of the Fe L emission, and its value indicates
that Fe 17$-$19 are the dominant Fe ions in this component.  The
constancy of the fitted ionization age with temperature shows that the
discrepancy between the Fe L and K intensities cannot be resolved with
a multi-temperature plasma with temperatures between 2.5 and 20 keV
unless a range of ionization ages is also invoked.

The additional Fe K emission evidently comes from a component that
does not contribute significantly to the Fe L emission, as the
spectrum is well-fit everywhere by a two-component model except at the
Fe K blend.  We consider two possibilities in the following sections:
Fe ejecta which are stratified interior to the other ejecta and
therefore at a different temperature and ionization age, and
fluorescent Fe K emission from Fe in dust grains, as was recently
proposed by Borkowski \& Szymkowiak (1997).

\subsection{Stratified Fe Ejecta}

In this section, we consider fit results for an ejecta component at
temperature $kT=$ 0.86 keV, a blast component at temperature $kT=$ 4 keV, and
an additional Fe emission component that we associate with ejecta.
The results of these fits are given in Table 3 for a range of
temperatures from 1$-$20 keV for the Fe ejecta.  Again, only the Si
and Fe abundances are given in the table for the 0.86 keV component,
but the abundances for the other elements relative to Si are similar
to those in the two-component fits of the previous section.

The lower limit of the Fe temperature is about twice the temperature
of the Si ejecta.  We do not attempt to constrain the upper limit of
the Fe ejecta temperature.  The Fe K centroid energy changes only
gradually with temperature above about $kT=$ 1.5 keV, so
good fits can be obtained at all these temperatures.  As the
temperature of the Fe ejecta is allowed to increase above the
temperature of the blast wave, the formal $\chi^2$ continues to
decrease gradually.  The ionization age of the Fe ejecta is roughly
100 times lower than the ionization age of the Si ejecta, so that the
dominant Fe ions are in low ionization stages: Fe 10$-$12, if their
temperature is about 1.6 keV, or Fe 14$-$17, if their temperature is
as high as 20 keV.  The low ionization age of the Fe ejecta component
suggests that the Fe ejecta were more recently shocked than the Si
ejecta, and had expanded to lower densities.  The exact translation
from an average ionization age to density is complicated, however,
because the ejecta are made of pure heavy elements and the electron
population evolves with time as the metals are gradually ionized.

The temperature of the blast wave in this situation is somewhat
uncertain.  We have fixed it to a value of 4.0 keV, which is between
the best fit temperature of 4.5 keV in the one-temperature NEI fits
and the best fit temperature of 3.6 keV in the two-temperature NEI
fits.  Because we have obtained a good fit for the parameters chosen,
and in light of the computational load required to search the entire
parameter space, we have not explored the full range of possible blast
wave temperatures.  This should be borne in mind.  Fits with blast
temperatures of 3.0, 3.5, or 4.5 keV were generally statistically
comparable to those with blast temperatures of 4.0 keV ($\Delta\chi^2$
within 2.7).

Figure 4 shows the model with the Fe ejecta at temperature $kT=$ 1.6
keV and the blast wave at temperature $kT=$ 4 keV.  As can be seen, the
fit is much improved over Figure 2, both formally, and to the eye.
From the figure, it is seen that the blast wave contributes most of
the emission in the Fe L region at energies between 0.7$-$1.0 keV.
The line emission makes up 56\% of the total count rate in the Fe K
band.  The blast wave component contributes 14\% of the observed count
rate in the Fe K line with an energy of 6.50 keV; the Fe ejecta at
temperature $kT=$ 1.6 keV contribute 70\% of the count rate in the line
with an energy of 6.41 keV.  We will address the remaining 16\% not
accounted for by the model in the discussion section.
%At the very low ionization age of the Fe ejecta,
%the major excitation process is innershell ionization.  The yield for
%L emission from the resulting K-shell vacancy is very low compared to
%that for K emission, which explains why this component of the model gives
%substantial Fe K emission with little accompanying Fe L emission.

\subsection{Dust Depletion and Fluorescence of Fe}

HG97 note that a possible alternative explanation for the low ratio of
Fe L to Fe K intensity is offered by Borkowski \& Szymkowiak (1997;
hereafter BS97), who point out that collisionally-excited Fe in dust
grains will fluoresce efficiently in the K transitions, but not the L
transitions.  If Fe is depleted onto dust grains, this results in a
low Fe L/Fe K intensity ratio: all the Fe line intensities are
decreased because the Fe is depleted onto dust, but the Fe K intensity
is supplemented by fluorescent emission from the Fe in dust grains,
while the Fe L intensity is not.  As noted by BS97, this is a
plausible scenario when dust is present in the remnant and the
ionization age is low enough that the observed Fe K line energy is
near 6.4 keV, the energy of the transition in neutral Fe.  Dust is
clearly present in Tycho, as is evidenced by its IR emission (Braun
1987).  The fitted energy centroid of the Fe K blend in data from the
best-calibrated ASCA CCD chip (SIS0, chip 1) is 6.471 (6.444$-$6.500,
$\Delta \chi^2$ = 2.71).  The systematic error in the line energy must
be taken to be its nominal value of 0.5\%, or about 0.03 keV for a
line energy of 6.4 keV.  There is unfortunately no independent way to
set the energy scale for Tycho's ASCA spectrum, as there are no strong
unblended lines at known energies.  Given the errors, the centroid of
the Fe K emission in Tycho is low enough that fluorescent K emission
from neutral Fe could be important.

We use the calculations of BS97 for the emissivity of a dusty gas of
Fe to calculate the mass of Fe required to reproduce Tycho's Fe K
flux.  The top panel of Figure 2 in BS97 plots the Fe K centroid as a
function of ionization age at the temperatures $kT=$ 1, 2, and 10 keV.
These calculations are carried out assuming (1) no dust and (2) an
initial depletion of 95\% of the Fe onto dust.  For temperatures other
than these, we interpolate between the calculations.  From the
measured Fe K centroid, the ionization age $n_et$ is $\sim 10^{10}$
cm$^{-3}$ s in gas without dust, which agrees well with the NEI fits
for the Fe K region in Table 1. The ionization age can be as high as
$n_et \sim 10^{10.3}$ cm$^{-3}$ s in gas with dust, depending on the
temperature.  Having set the ionization age to be $10^{10.3}$
cm$^{-3}$ s for the dusty gas, we then use the measured Fe K flux of
4.4 $\times 10^{-4}$ photons cm$^{-2}$ s$^{-1}$ from HG97 together
with emissivities plotted in Figure 1 of BS97 to deduce the Fe masses
given in Table 4.  These represent the total mass of Fe in both gas
and dust.  We can estimate how much of this mass is in gas by using
the fitted emission measures and abundances for the blast-shocked
component from our two-temperature fits (Table 2).  These masses are
also given in the table.  The implied mass of Fe in dust is between
0.002 M$_\odot$, if $kT$ = 10 keV for the blast wave, and 0.017
M$_\odot$, if $kT$ = 2.5 keV.  The calculations assume a spherical
shell of thickness 1/10 the radius and a distance of 2.3 kpc to
Tycho's SNR (Green 1984).

These mass calculations can be compared to the estimated mass in the
dust from the IRAS observations.  Braun (1987) estimates that the mass
of dust is 0.008 M$_\odot$ based on the IR brightness distribution and
intensity for a distance of 2.3 kpc.  For solar abundance ratios and a
mixture of graphite and silicate grains, about 0.002 M$_\odot$ of this
dust mass is Fe.  This is much lower than the dust masses in Table 4
at all temperatures below $kT=$ 10 keV.  However, processes involving
dust are complicated and all the calculations described have been
carried out with assumptions about the dust size distribution and
composition.  Uncertainties in these calculations and our application
of them are probably large enough to account for factors of a few, and
prevent us from ruling out this process as an important contribution
to the X-ray emission in Tycho.  Nevertheless it is clear that a
relatively large mass of dust is required if dust fluorescence is
important.

We have assumed that any dust present in the remnant is behind the
blast wave.  The presence of a significant amount of dust in the
ejecta may provide the mass of dust required by the X-ray data.
Certainly, the ASCA Fe K images suggest that the Fe K line emission is
distributed radially interior to the other emission lines, which would
support its origin in ejecta.  However, there are reasons to believe
that most of the IR-emitting dust in the remnant is from shocked
interstellar material.  The IR radial profiles peak at radii outside
the X-ray and radio radial profiles, which are both bright in emission
from the reverse shock through the ejecta (Braun 1987); this suggests
that most of the IR emission is from the blast wave.  Hamilton \&
Fesen (1988) argue that the relatively low ratio of X-ray to IR
luminosity in SN1006 implies that there cannot be a large quantity of
dust in the remnant; Tycho's SNR has a similar X/IR luminosity ratio,
so a similar argument would apply.  These considerations, in
combination with the expectation from the ASCA Fe K image that the Fe
K emission is primarily from the ejecta suggests that the dust
fluorescence scenario probably cannot be the primary explanation for
the discrepancy in Fe L/K intensities in the two-temperature models of
Section 2.1.  Nevertheless, since dust is present in Tycho's SNR, one
may expect that the process described here does occur.

\section{Discussion}

The ejecta plus blast emission model of Section 2.2 makes certain
predictions about the relative strength of the various spectral
components that can be compared to available imaging data.  Seward et
al. (1983) estimate that 28\% of the count rate at energies 0.5$-$4.5
keV in the Einstein High Resolution Imager (HRI) is due to emission
from the blast wave.  By folding our model through the Einstein HRI
response we find that a comparable fraction (32\%) of the counts in
the HRI band is due to the blast wave component.

We also consider the ASCA emission line images presented by HG97.  The
Fe K radial profile peaks at a smaller radius than the other ASCA
images, including the adjacent 4$-$6 keV continuum and the Fe L
emission at 1 keV.  This is evident in the radial profiles shown in
Figure 9 of HG97.  The Fe K image is made up of continuum emission
from the blast wave, plus line emission from both the blast wave and
the ejecta.  Since the 4$-$6 keV continuum image is dominated by the
blast wave emission, we assume that the continuum contribution to the
Fe K image has the same spatial distribution.  We can then subtract
the blast wave contribution to the Fe K image by scaling the 4$-$6 keV
radial profile to the Fe K radial profile at the outermost radii,
where the ejecta should make little or no contribution to the count
rate.  Subtracting this contribution from the total leaves about 32\%
of the Fe K image from the Fe ejecta.  Our spectral model with the Fe
ejecta at 4.0 keV predicts that this fraction is $\sim$ 40\%, which is
in reasonable agreement considering that these calculations are only
intended to be illustrative.  The spatial distribution of the Fe L
emission is consistent with its origin being predominatly in the blast
wave, as it is in our model.

The ejecta models for Tycho provide satisfactory fits overall, but
still slightly underestimate the Fe K flux by about 10\%.  The deficit
is comparable to the contribution of the blast wave to the Fe K line
emission.  Although we have argued that dust fluorescence is not the
dominant contribution to the Fe K line emission, it is certainly
possible that dust in the swept-up interstellar medium would
contribute an extra factor of two to the Fe K flux of the blast-wave
component.  The dust contribution would be at an energy of about 6.40
keV.  Depletion of interstellar material onto dust grains would also
explain why the fitted abundances in the blast wave component are
sub-solar.  It is also possible that a more sophisticated hydrodynamic
scenario is required; certainly all real remnants have complex
temperature, abundance, and density distributions that will manifest
themselves given sufficiently high quality data.

In our analysis, we have purposely focused on the data from the
best-calibrated chip of the SIS for technical reasons.  The
temperature of the gas in Tycho is certainly not uniform across the
face of the remnant, as the results of HG97 show.  However, the
temperature variations to which the ASCA data are sensitive are
relatively modest (roughly 50\%), and the reader is referred to HG97
for a discussion of the variation of the spectrum with position in the
remnant.

Our analysis indicates that the blast wave temperature $kT$ in Tycho's
SNR is from a few to several keV.  High energy flux has been detected
from the remnant by HEAO-1 (Pravdo \& Smith 1979) and Ginga (Fink et
al. 1994).  Fink et al. present spectral fits to the spectrum at
energies above 5 keV which require two components.  One is a thermal
component with temperature $kT$ about 3 keV, which is in agreement
with our blast wave component.  Fink et al. conclude that the second,
harder component could be either a hotter thermal component or a
nonthermal component, but could not constrain the parameters for
either.  If our analysis is correct, then it is suggestive that the
harder component measured by Fink et al. corresponds to nonthermal
X-ray emission such as was observed in SN1006 (Koyama et al. 1995).
If part of the continuum does come from such a nonthermal component,
the derived abundances of the elements must be increased, since
abundances are determined by the strength of the line emission
relative to the thermal continuum.  We have found that it is possible
to hide a nonthermal X-ray source underneath the brighter thermal
X-ray emission in the ASCA spectrum of Tycho with an unabsorbed
luminosity at energies 0.5$-$10 keV on the order of $10^{34}$ ergs
s$^{-1}$.  This is about an order of magnitude lower than the
luminosity of the nonthermal component in SN1006.  The nature of the
high energy emission from Tycho's SNR is an issue that will be of
great interest to address with the spectral instruments on the X-ray
Timing Explorer, which have sensitivity above 10 keV, where the
second, harder component of Fink et al.  begins to dominate.

We leave comparison of the data to a detailed hydrodynamical
simulation to others, but consider briefly what initial density
profiles for the ejecta may give rise to the observed temperature and
ionization age structure.  The power-law ejecta profiles described
with similarity solutions by Chevalier (1982) give precisely the
opposite trends in temperature and density as observed.  Assuming that
the Fe ejecta are interior to a mixed layer of lighter elements, the
initial power-law density profile in the ejecta would imply higher
density and lower temperature at inner radii.  The constant density
ejecta profiles described by Hamilton et al. (1986) do give a decrease
in density with decreasing radius, as required, but do not give an
increase in temperature because of the nonequilibration of the
electron temperature.  This model predicts an expansion rate of the
remnant that is higher than that measured in the radio, but consistent
with recent X-ray measurements (Hughes 1996).  A promising possibility
is the exponential ejecta density profile, which appears to be a
better description of Type Ia ejecta density profiles inferred from
the modelling of SN.  Although this model predicts a relatively flat
temperature profile and decreasing density profile, the presence of a
small amount of circumstellar matter may increase both the density and
temperature contrast (Dwarkadas \& Chevalier 1997).

\section{Summary and Conclusions}

A spectral model with three separate constant temperature and
ionization age NEI components provides a satisfactory fit to the ASCA
SIS spectrum.  This model makes no assumptions about the hydrodynamic
structure of the remnant.  One component representing ejecta is
included at the temperature $kT$ of 0.86 keV obtained by HG97 from
analysis of line intensity ratios.  The abundances of the intermediate
mass elements in this component are in good qualitative agreement with
those in Nomoto et al.'s (1984) W7 model for Type Ia supernovae.  The
low fitted abundance of Fe relative to Si suggests that much of the Fe
ejecta in Tycho's SNR are spatially separated from the Si ejecta.  A
homogeneous distribution is not possible given the large mass of Fe
ejecta expected in Type Ia remnants.  This is in accordance with
previously accumulated evidence (especially Hamilton, Sarazin, \&
Szymkowiak 1986) and with expectations from theory.  The model
therefore includes an additional component to represent Fe ejecta.
The blast wave is modelled as a third component with its element
abundances in the solar ratio.

The Fe L emission is relatively weak, and as noted, little of it is
associated with the Si and other ejecta.  The Fe K emission is
prominent, but does not come primarily from either the Si ejecta
component, which has too low a temperature, or the blast wave
component.  Most of the Fe K emission comes from Fe ejecta component,
at a temperature at least twice that of the Si ejecta, and an
ionization age about 100 times lower.  A slight residual deficiency in
the model Fe K flux may be explained by fluorescent emission from the
dust in the remnant, or possibly by the complex hydrodynamic structure
of the remnant.  The model is consistent with the currently available
imaging results.

Although fluorescent Fe K emission from dust can also qualitatively
explain the Fe L/Fe K intensity ratio and Fe K centroid energy in
Tycho, the mass of dust required to reproduce the Fe K intensity is
very high compared to the dust mass estimates from IRAS observations.
It cannot be ruled out, given the physical complexity of the process
(a detailed comparison to the data would be required), but we believe
that it is not likely to be the most important effect.

Our results leave some questions that will require more sophisticated
modelling for their resolution.  These include the slight residual
deficiency in the Fe K blend and the issues of dust depletion of the
elements in the interstellar medium and of a nonthermal X-ray
component in the spectrum.

\acknowledgments The authors thank Kazik Borkowski, Vikram Dwarkadas,
Eli Dwek, Eric Gotthelf, and Andy Szymkowiak for profitable scientific
discussions.  We also express appreciation for the late Tom Markert,
who provided input and encouragement.  UH acknowledges support from an
NRC resesarch associateship.  JPH acknowledges support from NASA grant
NAG 5-4794.

\clearpage
\begin{deluxetable}{ccccc}
\tablecaption{NEI Fits for E $>$ 4.5 keV}
\tablewidth{0pt}
\tablehead{
\colhead{Blast $kT$} &\colhead{$\chi^2$} & \colhead{Fe\tablenotemark{a}}
& \colhead{log $\tau$} & \colhead{EM\tablenotemark{b}} \nl
\colhead{(keV)} &\colhead{($\nu$=27)} & \colhead{(rel $\odot$)} &
\colhead{(cm$^{-3}$ s)} & \colhead{($10^{12}$ cm$^{-5}$)} \nl
}
\startdata
2.5  & 22.0 & 3.4 (1.9$-$8.4) & 10.12 & 8.3 \nl
3.0  & 19.4 & 3.1 (1.8$-$6.7) & 10.07 & 6.1 \nl
3.3  & 18.0 & 2.9 (1.8$-$5.3) & 10.06 & 5.0 \nl
4.0  & 17.2 & 2.6 (1.6$-$4.8) & 10.03 & 3.9 \nl  
4.5  & 16.9 & 2.5 (1.5$-$4.2) & 10.03 & 3.5 \nl
5.0  & 17.1 & 2.4 (1.4$-$4.0) & 10.02 & 3.1 \nl  
6.0  & 17.4 & 2.2 (1.4$-$3.6) & 10.02 & 2.7 \nl
7.0  & 18.3 & 2.1 (1.3$-$3.4) & 10.02 & 2.4 \nl  
10.0 & 20.3 & 1.9 (1.2$-$2.9) & 10.03 & 2.0 \nl         
20.0 & 24.2 & 1.6 (1.0$-$2.3) & 10.10 & 1.8 \nl
\enddata
\tablenotetext{a}{Abundances are relative to assumed solar value of
3.98 $\times 10^{-5}$ relative to H by number (Allen 1973).  The range
in parentheses corresponds to $\Delta\chi^2$ = 2.71.}
\tablenotetext{b}{Emission measures should be scaled by $\sim$ 2.3 to
normalize to the entire remnant.}
\end{deluxetable}

\clearpage
\psfig{figure=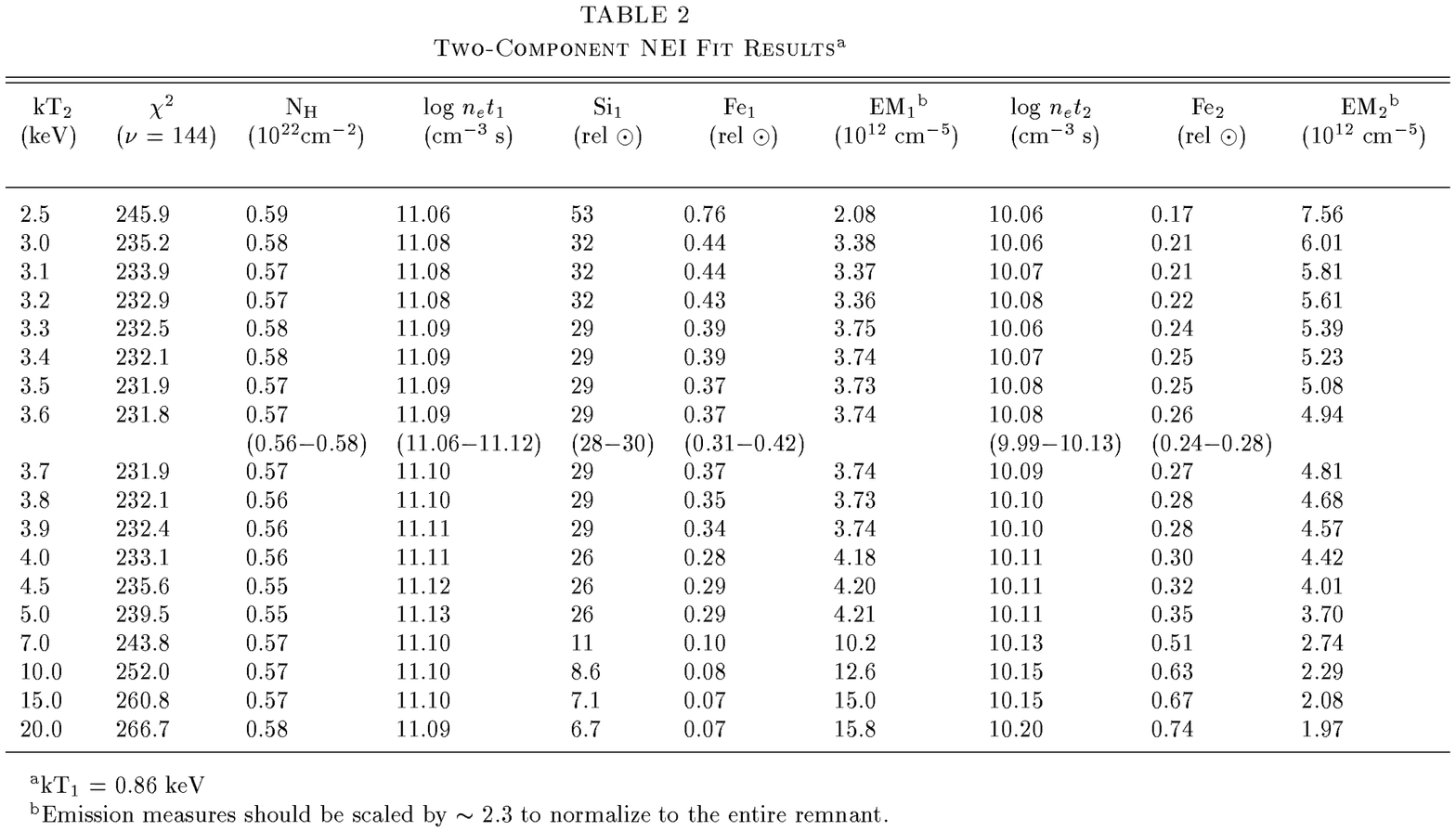,width=6.5truein,angle=90.0}

\clearpage
\psfig{figure=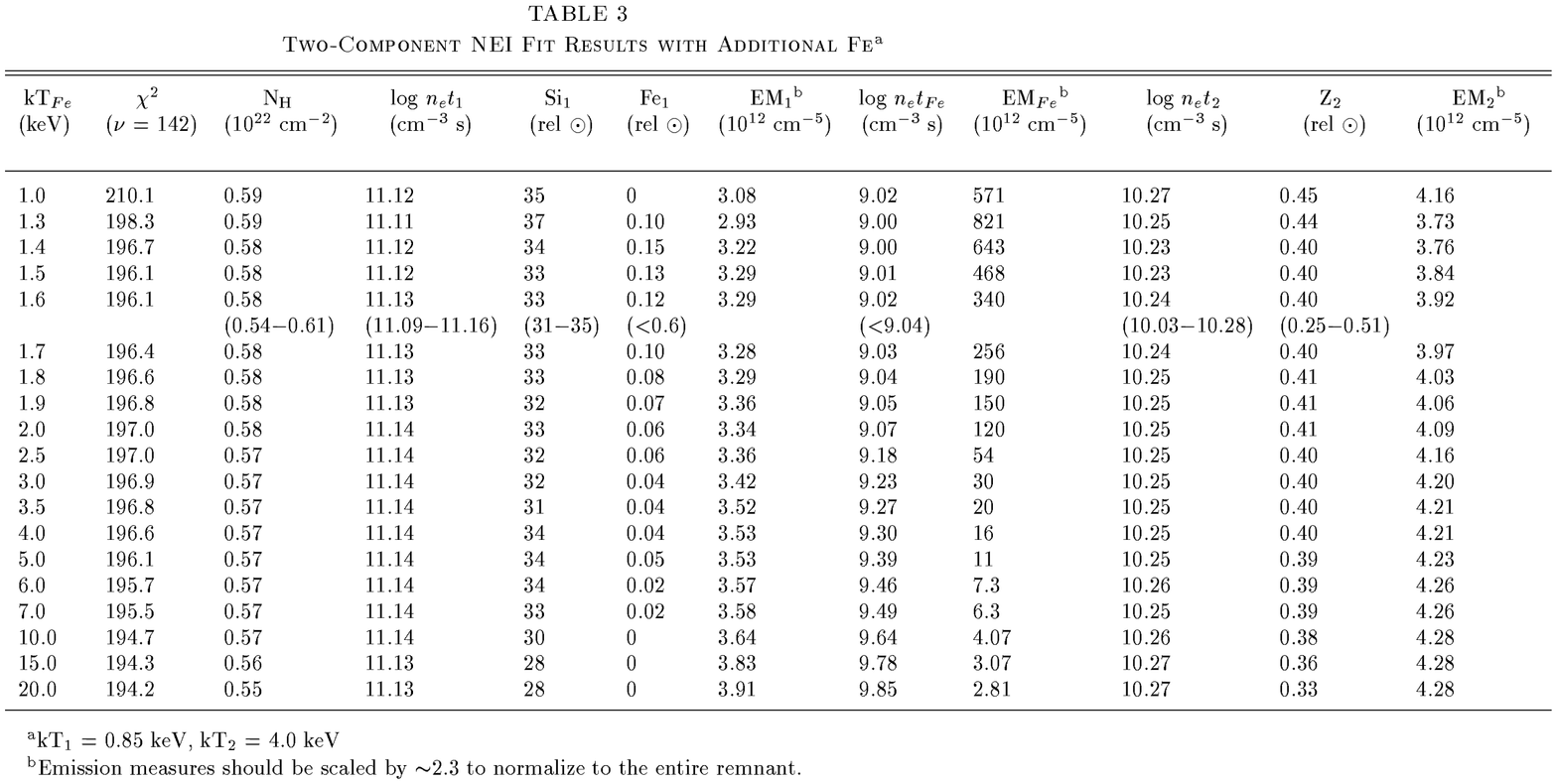,width=6.5truein,angle=90.0}

\clearpage
\setcounter{table}{3}
\begin{deluxetable}{cccc}
\tablecaption{Fe Masses}
\tablewidth{0pt}
\tablehead{
\colhead{Blast $kT$} &\colhead{M$_{\rm Fe}$ (\msun)}         &\colhead{M$_{\rm Fe}$ (\msun)}&\colhead{M$_{Fe}$ (\msun)} \nl
\colhead{(keV)}    &\colhead{(gas + dust)\tablenotemark{a}}&\colhead{gas}                 &\colhead{dust}\nl
}
\startdata
2.5  &  0.018  & 0.0011 & 0.017 \nl
3.0  &  0.013  & 0.0012 & 0.012 \nl
3.3  &  0.011  & 0.0012 & 0.010 \nl
4.0  &  0.009  & 0.0014 & 0.007 \nl
5.0  &  0.007  & 0.0015 & 0.006 \nl
6.0  &  0.006  & 0.0017 & 0.005 \nl
7.0  &  0.005  & 0.0017 & 0.004 \nl
10.0 &  0.004  & 0.0017 & 0.002 \nl
\enddata
\tablenotetext{a}{Using BS97 for log $n_et$ (cm$^{-3}$ s) = 10.3 }
\end{deluxetable}

\begin{figure}
\centerline{
{\hfil\hfil
\psfig{figure=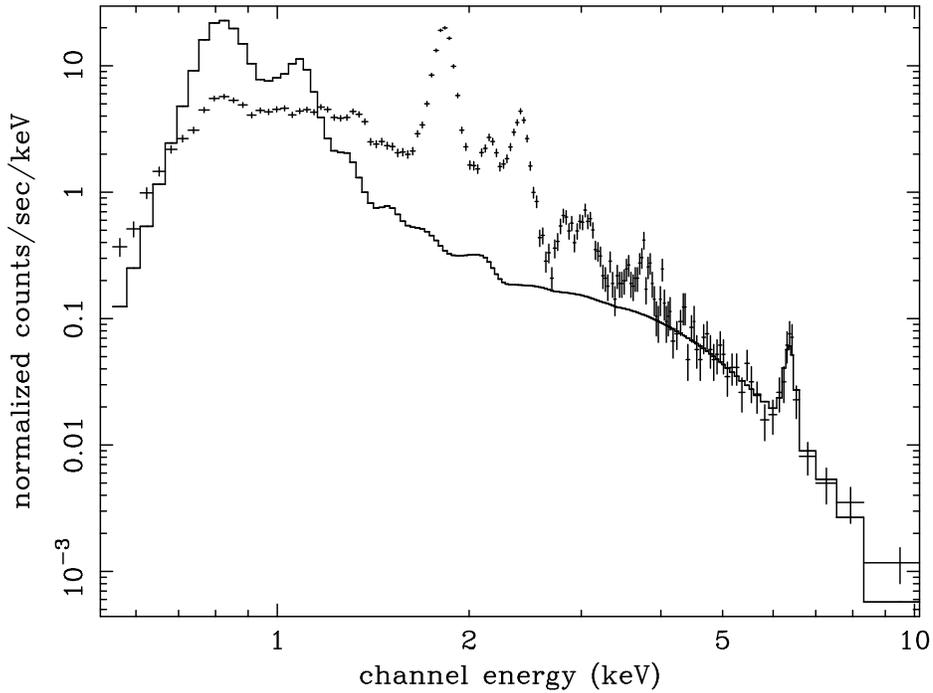,height=4.0truein,angle=270.0,clip=}
\hfil\hfil}}
\caption{The ASCA (SIS0, chip 1) spectrum of Tycho's SNR overlaid
with the best-fit NEI model for continuum and Fe, fitted at energies
above 4.5 keV and extrapolated to lower energies.  The Fe L emission
is severely overpredicted, even in the absence of the ejecta component
required to explain the strong Si and S line emission at 1.86 and 2.45
keV.  The model assumes the Galactic column density of $4.5 \times
10^{21}\ \rm{cm}^{-2}$ and is folded through the instrument response.}
\end{figure}

\begin{figure}
\centerline{
{\hfil\hfil
}}
\centerline{
{\hfil\hfil
\psfig{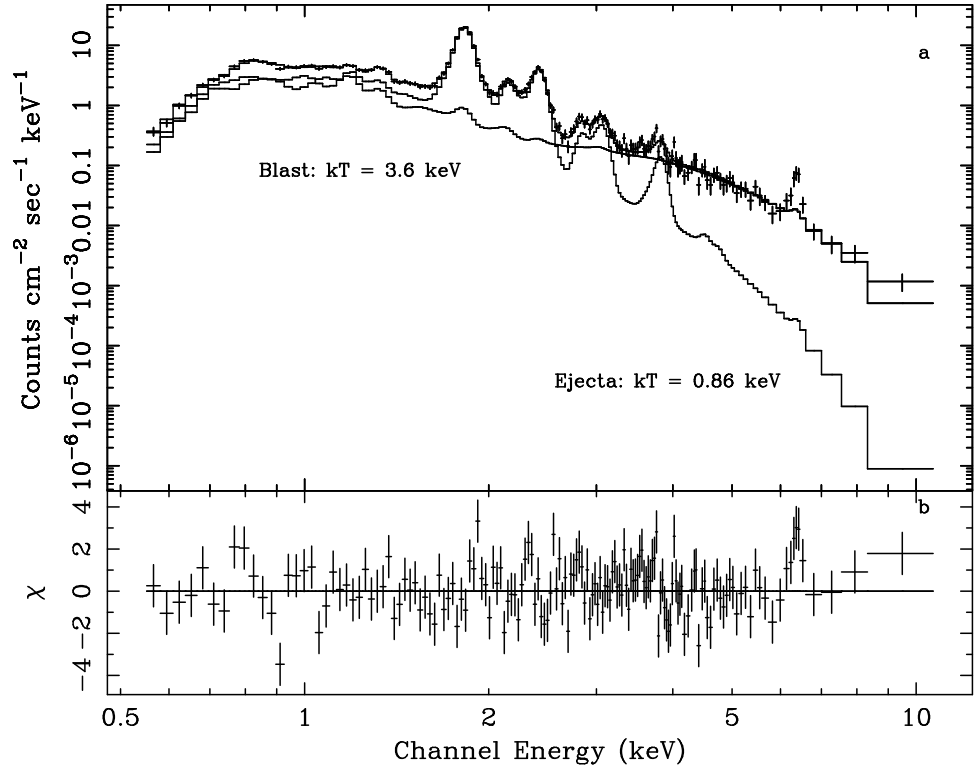}
}}
\caption{The data overlaid with the two temperature NEI model with
ejecta temperature $kT$ = 0.86 keV and blast temperature $kT$ = 3.6
keV (the model with the lowest formal $\chi^2$).  The lower panel
shows the residuals.  The contribution of each component is shown
separately.  The ejecta component can be recognized as it contributes
most of the strong Si and S emission, while the blast component
provides most of the continuum at energies above 5 keV.  This model,
constrained by the Fe L emission, provides insufficient emission in
the Fe K blend at $\sim$6.5 keV.}
\end{figure}

\begin{figure}
\centerline{
{\hfil\hfil
\psfig{figure=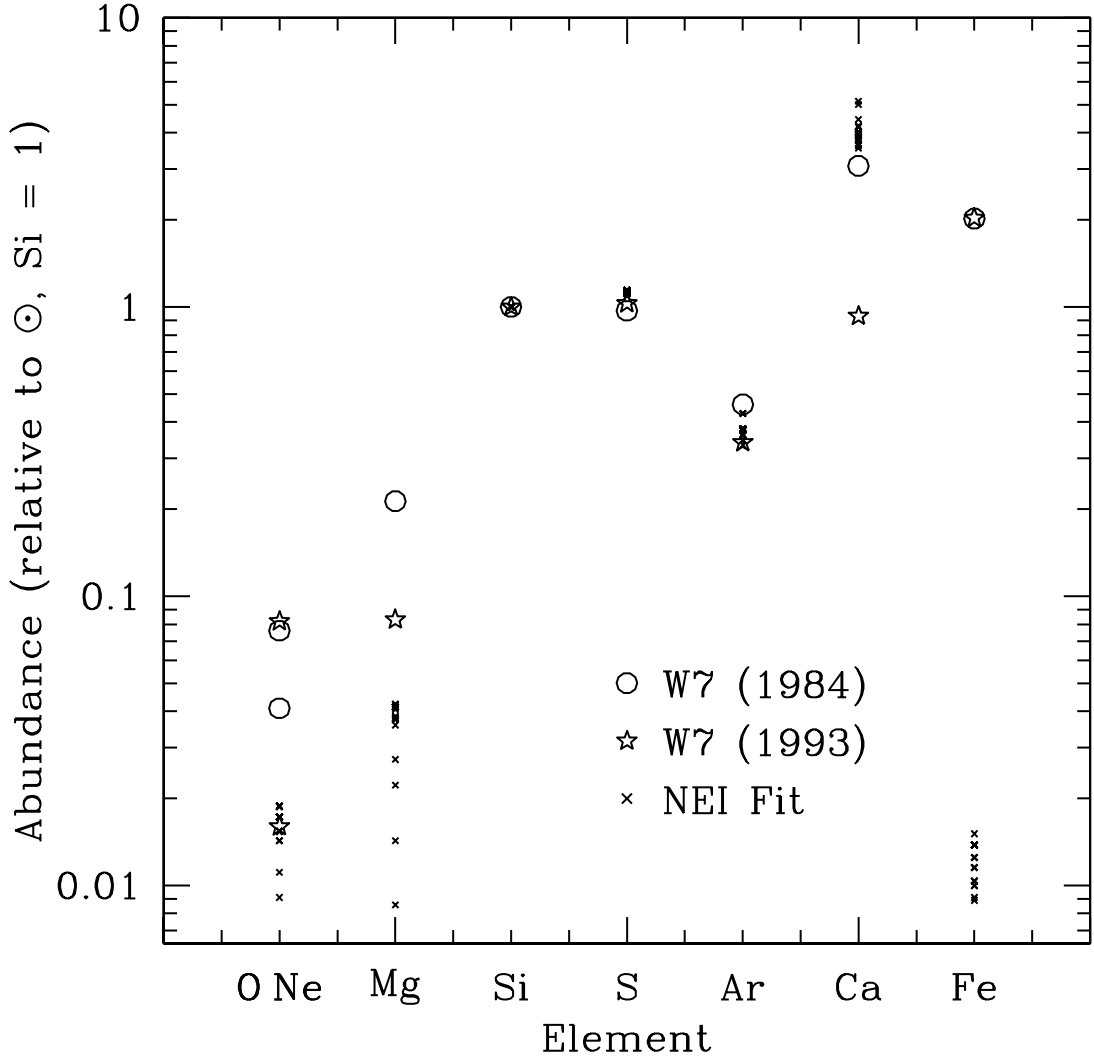,height=6.0truein,angle=0,clip=}
\hfil\hfil}
}
\caption{The abundances of the elements relative to Si, relative to
the solar values of Allen (1973) from the fits in Table 2.  Also shown
are the abundances calculated for the W7 model for a Type Ia supernova
by Nomoto et al. (1984) and the updated calculations by Thielemann et
al. (1993).  O and Ne are shown in the same column; in both models,
the Ne abundance is lower than the O abundance. }
\end{figure}

\begin{figure}
\centerline{
{\hfil\hfil
\hfil\hfil}}
\centerline{
{\hfil\hfil
\psfig{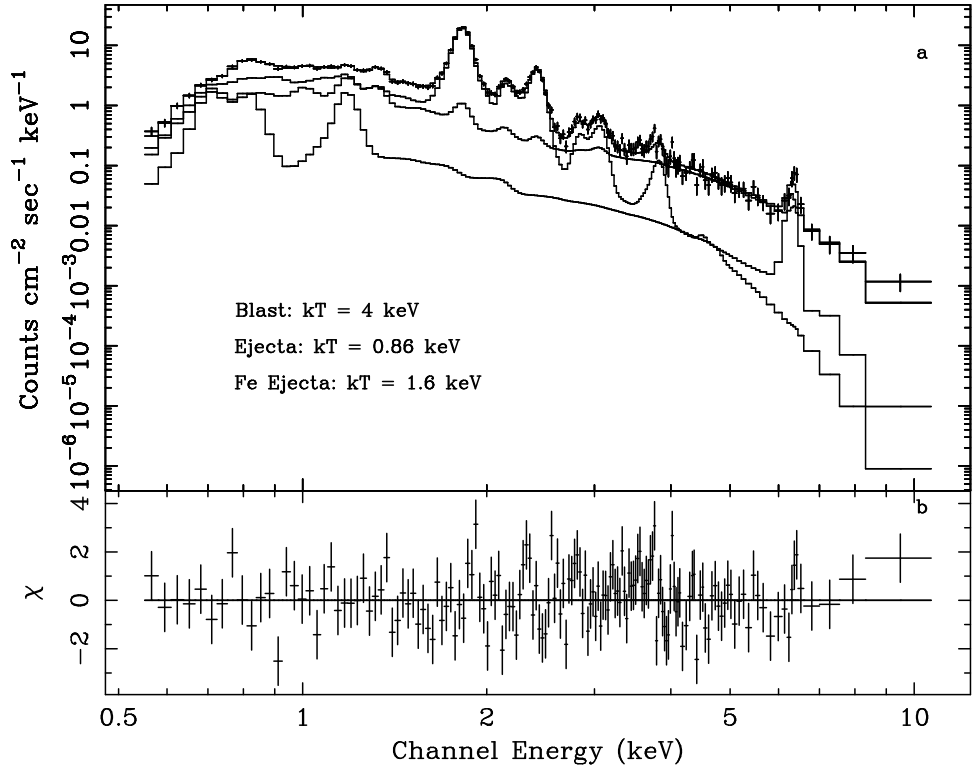}
\hfil\hfil}}
\caption{The data overlaid with the NEI model with ejecta at
temperature $kT =$ 0.86 keV, additional Fe ejecta at temperature $kT=$
1.6 keV, and blast component at temperature $kT=$ 4 keV, folded through
the instrument response.  The lower panel shows the residuals. The
contribution of each component is shown separately.  The ejecta
component can be recognized as it contributes most of the strong Si
and S emission, while the blast component provides most of the
continuum at energies above 5 keV.  The curves for the ejecta and
blast wave cross at about 1.2 keV. }
\end{figure}

\end{document}